\documentclass[journal,a4paper,10pt]{IEEEtran}
\usepackage[T1]{fontenc}
\usepackage[latin9]{inputenc}

\usepackage{epsfig,psfrag}
\usepackage{amsmath,amsfonts,amssymb,amsxtra,bm}
\usepackage{graphicx}
\usepackage{color}
\usepackage{cite}

\newcommand{\eq}{\,=\,}
\newcommand{\be}{\begin{equation}}
\newcommand{\ee}{\end{equation}}
\newcommand{\ist}{\hspace*{.3mm}}
\newcommand{\rmv}{\hspace*{-.3mm}}

\definecolor{hl}{rgb}{0,0,1}

\allowdisplaybreaks

\begin{document}

\title{Simultaneous Distributed Sensor Self-Localization and Target Tracking Using Belief Propagation and Likelihood Consensus\vspace{2mm}}
\author{\emph{Florian Meyer\hspace{-.05mm}, Erwin Riegler\hspace{-.05mm}, Ondrej Hlinka\hspace{.15mm}, and Franz Hlawatsch}\\[2.5mm]
{\normalsize Institute of Telecommunications, Vienna University of Technology, Austria (florian.meyer@tuwien.ac.at)}
\thanks{This work was supported by the Austrian Science Fund (FWF) 
under Award S10603 (Statistical Inference) within the National Research Network SISE
  and by the WWTF under Award ICT10-066 (NOWIRE).}\vspace{0mm}}
\maketitle

\thispagestyle{empty}
\pagestyle{empty}

\begin{abstract}
We introduce the framework of 
\emph{cooperative simultaneous localization and tracking} (CoSLAT), which provides a 
%% complete and 
consistent combination of cooperative self-localization (CSL) and distributed target tracking (DTT)
in sensor networks without a fusion center.
CoSLAT extends 
%% the recently introduced 
simultaneous localization and tracking (SLAT) 
%% framework 
in that it 
%% allows for mobile sensors and 
uses also intersensor measurements. 
Starting from a factor graph formulation of the CoSLAT problem, we develop a particle-based, 
distributed message passing algorithm for CoSLAT
that combines nonparametric belief propagation with the 
%% the recently introduced 
likelihood consensus scheme.
The proposed CoSLAT algorithm improves on state-of-the-art CSL and DTT algorithms by 
exchanging probabilistic information between CSL and DTT.
%%  that is allowed by the CoSLAT framework. The proposed CoSLAT algorithm 
%% integrates DTT into nonparametric belief propagation for CSL. It 
%% The latter is used to convey relevant information 
%% about the target position 
%% available 
%% to all sensors in a distributed way. 
Simulation results demonstrate substantial improvements in both self-localization 
%% performance 
and tracking performance. 
%% achieved with the CoSLAT algorithm in comparison to 
%% compared to 
%% separate 
%% state-of-the-art 
%% CSL and DTT 
%% algorithms. 
%% These improvements are due to an exchange of probabilistic information between CSL and DTT.
\vspace{.1mm}
\end{abstract}

\begin{keywords}
%% Wireless sensor network, 
Distributed target tracking, cooperative localization, CoSLAT, nonparametric belief propagation, likelihood consensus.
\vspace{-3mm}
\end{keywords}

%%%%%%%%%%%%%%%%%%%%%%%%%%%%%%%%%%%
\section{Introduction}\label{sec:intro}
%%%%%%%%%%%%%%%%%%%%%%%%%%%%%%%%%%%

Two important 
%% distributed 
inference tasks in decentralized sensor networks are cooperative self-localization (CSL) \cite{patwari,wymeersch}
and distributed target tracking (DTT) \cite{liu07}. 
In CSL, each sensor acquires measurements of its own location relative to neighboring sensors,
and it cooperates with all the other sensors to estimate its own location.
Existing CSL algorithms include\linebreak %%%%%%%% 
nonparametric belief propagation (NBP) \cite{ihler} and other message passing algorithms \cite{wymeersch,pedersen}.
In DTT, each sensor acquires a measurement that is related to the state of a target, and it cooperatively estimates 
the target state based on the measurements of all sensors. 
Existing DTT algorithms include consensus-based distributed particle filters \cite{hlinka,farahmand,savic}.
In the framework of distributed \emph{simultaneous localization and tracking} (SLAT) \cite{taylor}, the sensors simultaneously track a target and 
localize themselves, however without using intersensor distance measurements. 
Methods for SLAT were proposed in \cite{taylor, funiak, garciafernandez, oguzekim, chen11, kantas12}. 

CSL and DTT are closely related since (i) to contribute to DTT, a sensor needs to have information of its own location, and (ii) 
the accuracy of CSL may be improved if the sensors possess estimates of the 
%% (location-related) 
state of a target.
This observation motivates the development of combined CSL-DTT methods.

Here, we introduce the framework of \emph{cooperative simultaneous localization and tracking} (CoSLAT), which,
for the first time, provides a consistent combination of CSL and DTT.
%% In CoSLAT, a target is tracked while simultaneously localizing the sensors, using measurements of the distances between the sensors and the target 
%% as well as measurements of the distances between the sensors.
%% Thus, 
CoSLAT extends SLAT in that it uses also intersensor distance measurements.
%%  and the sensors may be mobile.
%%  between the sensors. 
%% As we will demonstrate, this leads to improved performance of both sensor localization and target tracking.
%% ; however, the associated factor graph is more complex than in the case of SLAT. 
%% Our main contribution is a
%% , Bayesian, 
We propose a particle-based, distributed CoSLAT algorithm that integrates
%% includes 
DTT in NBP-based CSL \cite{ihler,wymeersch,lien}.
%% The proposed algorithm is, to the best of our knowledge, the first method for simultaneous CSL and DTT, and it
%% It takes full advantage of the exchange of probabilistic information between CSL and DTT that is allowed by the CoSLAT framework.
%% The 
%% non-cooperative 
%% targets are noncooperative in that they do not acquire any measurements nor do they communicate with other sensors. 
%% However, some of the sensors possess noisy measurements of their distance to the targets. 
%% Since the targets cannot perform any calculations, each 
%% Each sensor estimates both the target location (or, more generally, the target state) and its own location, using a BP message passing algorithm derived from
%% a factor graph formulation \cite{loeliger}. 
A fundamental problem---the
%% that has to be surmounted in this scheme is that
%% , as in pure DTT, 
nonavailability of essential information 
%% fact that information needed to calculate certain target-related messages is not available 
%% locally 
at the sensors---is 
%% The proposed CoSLAT algorithm 
%% combines a BP message passing algorithm based on a factor graph formulation \cite{loeliger} with 
%% uses 
solved by using the likelihood consensus (LC) scheme \cite{hlinka, hlinka2}.
%%  for a consensus-based distributed calculation of the lacking messages.
%% This results in a coherent combination of BP with LC.
%%  which, possibly, may also be of interest in other problems.
%%  besides the CoSLAT problem studied in this paper.
The algorithm's main new feature 
%% of the proposed CoSLAT algorithm 
is a probabilistic information transfer between CSL and DTT,
which allows CSL and DTT to support each other. As we will demonstrate, this leads to improved performance of both sensor localization and target tracking.

This paper is organized as follows. 
The system model is described in Section \ref{sec:wsn}.  
In Section \ref{sec:slat}, the CoSLAT problem is defined and a basic message passing scheme for CoSLAT is derived.
%% the novel CoSLAT algorithm is presented. 
This scheme is further developed into a 
%% practical 
distributed CoSLAT algorithm in Section \ref{sec:slat_practical}.
Finally, 
%% the performance of our CoSLAT algorithm is studied via 
simulation results are presented in Section \ref{sec:simres}.
%%  demonstrate the improvements in both self-localization performance and tracking performance achieved with the CoSLAT algorithm 
%% in comparison to separate CSL and DTT.

\begin{figure}[b!]
\vspace*{-2mm}
\psfrag{S1}[l][l][.8]{\raisebox{-1mm}{\hspace{-2mm}sensor}}
\psfrag{S2}[l][l][.8]{\raisebox{-2.5mm}{\hspace{-2mm}target}}
\psfrag{S3}[l][l][.8]{\raisebox{1mm}{\hspace{-2mm}communication link}}
\psfrag{S4}[l][l][.8]{\raisebox{0.5mm}{\hspace{-2mm}measurement link}}
\hspace{5mm}
\includegraphics[scale=0.52]{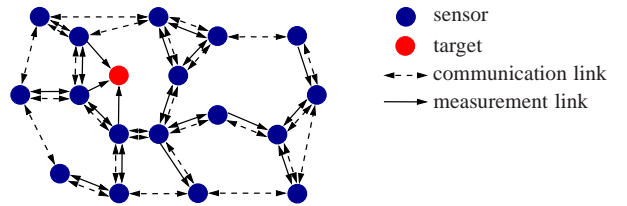}
\renewcommand{\baselinestretch}{1}\small\normalsize
\vspace*{-1mm}
\caption{Wireless sensor network with target, communication links, and mea\-sure\-ment links.}
\label{fig:wsn}
%% \vspace*{-1mm}
\end{figure}

\vspace{-1mm}

%%%%%%%%%%%%%%%%%%%%%%%%%%%%%%%%%%%
\section{System Model}\label{sec:wsn}
%%%%%%%%%%%%%%%%%%%%%%%%%%%%%%%%%%%

\vspace{.5mm}

We consider a sensor network consisting of $K$ cooperating sensor 
%% (CS) 
nodes and a noncooperative target node,
as depicted in Fig.\ \ref{fig:wsn}. The set of all nodes is $\mathcal{A} = \{0,\dots,K\}$, with $k = 0$ indexing 
the 
%% noncooperative 
target and $k \in \mathcal{A}_{\sim 0} \triangleq \mathcal{A} \backslash \{0\}$ indexing the 
%% cooperating 
sensors. 
%% The sensors cooperate in that they may communicate and may perform measurements and calculations. 
Sensors and target may be mobile. The \emph{state} of sensor or target $k \in \mathcal{A}$ at time $n \in \{0,1,\ldots\}$, 
denoted by $\mathbf{x}_{k,n}$, consists of the current location and, possibly, additional motion parameters such as velocity
%% some motion information such as the current velocity, 
%% acceleration, and/or angular velocity 
\cite{rong}.  The states $\mathbf{x}_{k,n}$
%% , $k \in \mathcal{A}$ 
evolve according to the state transition probability density functions (pdfs) $f(\mathbf{x}_{k,n}|\mathbf{x}_{k,n-1})$ and the state priors $f(\mathbf{x}_{k,0})$.

%% \newpage %%%%%%%%

The communication and measurement topologies 
%% of the sensor network 
are described by sets $\mathcal{C}_{n}$, $\mathcal{M}_{k,n}$, and $\mathcal{T}_n$ as follows. Two 
%% different 
sensors $k, l \rmv\in\rmv \mathcal{A}_{\sim 0}$ are able to communicate with each other if 
$(k,l) \rmv\in\rmv \mathcal{C}_{n} \rmv\subseteq\rmv \mathcal{A}_{\sim 0} \!\times\! \mathcal{A}_{\sim 0}$. $\mathcal{C}_{n}$ is symmetric, i.e., 
if $(k,l) \rmv\in\rmv \mathcal{C}_{n}$ then $(l,k) \rmv\in\rmv \mathcal{C}_{n}$.
%% which is true if their distance is not larger than a given communication radius $C$. 
Sensor $k \!\in\! \mathcal{A}_{\sim 0}$ acquires a measurement $y_{k,l;n}$\linebreak %%%%%%%
relative to sensor $l \in \mathcal{A}_{\sim 0}$, with $(k,l) \rmv\in\rmv \mathcal{C}_{n}$, 
if $l \in \mathcal{M}_{k,n} \rmv\subseteq \mathcal{A} \setminus\rmv \{k\}$.
Sensor $k \!\in\! \mathcal{A}_{\sim 0}$ acquires a measurement $y_{k,0;n}$ relative\linebreak %%%%%%% 
to the target, i.e., $0 \rmv\in\rmv \mathcal{M}_{k,n}$,
if $k \in \mathcal{T}_n \subseteq \mathcal{A}_{\sim 0}$; i.e., $\mathcal{T}_n \triangleq$\linebreak %%%%%%%%  
$\{ k\rmv\in\rmv \mathcal{A}_{\sim 0} \ist |\ist  0 \rmv\in\rmv \mathcal{M}_{k,n}\}$.
The sets $\mathcal{C}_{n}$, $\mathcal{M}_{k,n}$, and $\mathcal{T}_n$ may be\linebreak %%%%%%%% 
time-dependent.
%% ; note that $\mathcal{T}_n$ is implied by $\mathcal{M}_{k,n}$.
An example of communication and measurement topologies is given in Fig.\ \ref{fig:wsn}.
%% ???In this paper, we 
We consider a two-dimen\-sional (2D) scenario and noisy distance measurements 
%% modeled as
\begin{equation}
\label{eq:mess}
y_{k,l;n} \eq \| \tilde{\mathbf{x}}_{k,n} \!-\rmv \tilde{\mathbf{x}}_{l,n} \| \ist+\ist v_{k,l;n} \,,
\end{equation}
where $\tilde{\mathbf{x}}_{k,n} \!\triangleq\rmv [x_{1,k,n}\,\,\ist x_{2,k,n}]^\text{T}\rmv$ represents the location of sensor or target $k$
(note that this a part of the state $\mathbf{x}_{k,n}$). 
The measurement noise $v_{k,l;n}$ is not necessarily Gaussian; its variance $\sigma^2_v$ is assumed known; 
and $v_{k,l;n}$ and $v_{k'\!,l';n'}$ are assumed independent unless $(k,l,n)=(k'\!,l'\!,n')$.
We note that other measurement models could be used, and the extension to the 3D case is straightforward.

%% \newpage %%%%%%%

\vspace{-1mm}

%%%%%%%%%%%%%%%%%%%%%%%%%%%%%%%%%%%
\section{A Message Passing Scheme}\label{sec:slat}  %% CoSLAT, Part I:\, 
%%%%%%%%%%%%%%%%%%%%%%%%%%%%%%%%%%%

\vspace*{.5mm}

We first define the CoSLAT problem and derive a 
%% basic 
message passing scheme for CoSLAT.
This scheme will be 
%% further 
developed into a 
%% practical 
distributed CoSLAT algorithm in Section \ref{sec:slat_practical}.

In CoSLAT, at time $n$, each sensor $k \rmv\in\rmv \mathcal{A}_{\sim 0}$ estimates both\linebreak %%%%%%% 
its own state $\mathbf{x}_{k,n}$ and the target state $\mathbf{x}_{0,n}$,
using all the inter\-sensor and sensor-target distance measurements up to time $n$, i.e., 
%% This estimation is based on the \emph{entire measurement set} 
$\mathcal{Y}_{1:n} \triangleq {\{ y_{k,l;n'} \rmv \}}_{k \in \mathcal{A}_{\sim 0}, \,l \in \mathcal{M}_{k,n'}, \,n'\in \{1,\ldots,n\}}$.
\vspace*{.3mm}
In particular, the minimum mean square error (MMSE) estimator \cite{kay} of state $\mathbf{x}_{k,n}$
%% , $k \!\in\! \mathcal{A}$ 
%% based on $\mathcal{Y}_{1:n}$ 
is given by
\begin{equation}
\label{eq:mmse_coslat}
\hat{\mathbf{x}}^\text{MMSE}_{k,n} \ist\,\triangleq\,\ist \text{E}\{\mathbf{x}_{k,n}|\mathcal{Y}_{1:n}\} 
  \,= \int \rmv \mathbf{x}_{k,n} \ist f(\mathbf{x}_{k,n}|\mathcal{Y}_{1:n}) \ist d\mathbf{x}_{k,n} \,, 
\end{equation}
for all $k \!\in\! \mathcal{A}$. 
%% It is important to note that 
Compared to ``pure CSL'' \cite{wymeersch, ihler, lien, pedersen} 
and ``pure DTT'' \cite{farahmand,hlinka,savic}, 
the measurement set $\mathcal{Y}_{1:n}$ 
%% from which these state estimates are derived 
is extended in that it includes also the respective other measurements 
(i.e., sensor-target distance measurements for the sensor state estimates $\hat{\mathbf{x}}^\text{MMSE}_{k,n}$, $k \!\in\! \mathcal{A}_{\sim 0}$ 
and intersensor distance measurements for the target state estimate $\hat{\mathbf{x}}^\text{MMSE}_{0,n}$).
%% This is the reason why CoSLAT tends to outperform separate CSL and DTT as well as SLAT (which does not use the inter-sensor distance measurements).

%% \newpage %%%%%%%%.69]

\begin{figure}
\vspace*{2.1mm}
\centering
\psfrag{S17}[l][l][.8]{\raisebox{-2.2mm}{\hspace{-0.7mm}$\mathbf{x}_{1}$}}
\psfrag{S3}[l][l][.8]{\raisebox{-2mm}{\hspace{-0.7mm}$\mathbf{x}_{1}$}}
\psfrag{S21}[l][l][.8]{\raisebox{-2mm}{\hspace{-0.9mm}$\mathbf{x}_{2}$}}
\psfrag{S24}[l][l][.8]{\raisebox{-1.3mm}{\hspace{-1.4mm}$\mathbf{x}_{K}$}}
\psfrag{S7}[l][l][.8]{\raisebox{-2.2mm}{\hspace{-1.2mm}$\mathbf{x}_{2}$}}
\psfrag{S12}[l][l][.8]{\raisebox{-2mm}{\hspace{-1.6mm}$\mathbf{x}_{K}$}}
\psfrag{S2}[l][l][.7]{\raisebox{-1mm}{\hspace{-0.7mm}$f_1$}}
\psfrag{S18}[l][l][.7]{\raisebox{-1mm}{\hspace{-0.7mm}$f_1$}}
\psfrag{S6}[l][l][.7]{\raisebox{-1mm}{\hspace{-0.7mm}$f_2$}}
\psfrag{S20}[l][l][.7]{\raisebox{-1mm}{\hspace{-0.7mm}$f_2$}}
\psfrag{S13}[l][l][.7]{\raisebox{-1mm}{\hspace{-1.3mm}$f_K$}}
\psfrag{S23}[l][l][.7]{\raisebox{-3mm}{\hspace{-1mm}$f_K$}}
\psfrag{S5}[l][l][.7]{\raisebox{1.5mm}{\hspace{0.3mm}$\tilde{f}_1$}}
\psfrag{S8}[l][l][.7]{\raisebox{1.5mm}{\hspace{0.3mm}$\tilde{f}_2$}}
\psfrag{S11}[l][l][.7]{\raisebox{1.5mm}{\hspace{-0.3mm}$\tilde{f}_K$}}
\psfrag{S16}[l][l][.8]{\raisebox{0mm}{\hspace{-3mm}$n \!-\! 1$}}
\psfrag{S15}[l][l][.8]{\raisebox{0mm}{\hspace{1mm}$n$}}	
\psfrag{S4}[l][l][.7]{\raisebox{-3.7mm}{\hspace{-2.1mm}$f_{1,2}$}}
\psfrag{S19}[l][l][.7]{\raisebox{-3mm}{\hspace{-2.1mm}$f_{1,2}$}}
\psfrag{S31}[l][l][.7]{\raisebox{-3mm}{\hspace{-1mm}$f_{2,1}$}}
\psfrag{S33}[l][l][.7]{\raisebox{-3mm}{\hspace{-1mm}$f_{2,1}$}}
\psfrag{S32}[l][l][.7]{\raisebox{-3.5mm}{\hspace{-2mm}$f_{1,k}$}}
\psfrag{S34}[l][l][.7]{\raisebox{-4mm}{\hspace{-1.3mm}$f_{1,k}$}}
\psfrag{S35}[l][l][.7]{\raisebox{-2.5mm}{\hspace{-1.8mm}$f_{K,k}$}}
\psfrag{S36}[l][l][.7]{\raisebox{-2.5mm}{\hspace{-2mm}$f_{K,k}$}}
\psfrag{S9}[l][l][.7]{\raisebox{0mm}{\hspace{-1mm}$f_{2,k}$}}
\psfrag{S22}[l][l][.7]{\raisebox{0mm}{\hspace{-1mm}$f_{2,k}$}}
\psfrag{S28}[l][l][.69]{\raisebox{2mm}{\hspace{-1.7mm}\textcolor{hl}{$b_1^{(P)}$}}}
\psfrag{S29}[l][l][.69]{\raisebox{2mm}{\hspace{-2.9mm}\textcolor{hl}{$b_2^{(p-1)}$}}}
\psfrag{S30}[l][l][.69]{\raisebox{1mm}{\hspace{-3.5mm}\textcolor{hl}{$b_{k}^{(p-1)}$}}}
\psfrag{S26}[l][l][.69]{\raisebox{2.7mm}{\hspace{-4.5mm}\textcolor{hl}{$m_{k\rightarrow 1}^{(p)}$}}}
\psfrag{S25}[l][l][.69]{\raisebox{1mm}{\hspace{-1.7mm}\textcolor{hl}{$m_{2\rightarrow 1}^{(p)}$}}}
\psfrag{S27}[l][l][.69]{\raisebox{0mm}{\hspace{-2.5mm}\textcolor{hl}{$m_{\rightarrow n}$}}}
\psfrag{S45}[l][l][.8]{\raisebox{-2.3mm}{\hspace{-0.6mm}$\mathbf{x}_{0}$}}
\psfrag{S46}[l][l][.8]{\raisebox{-2.4mm}{\hspace{-0.6mm}$\mathbf{x}_{0}$}}
\psfrag{S43}[l][l][.7]{\raisebox{-2mm}{\hspace{-0.1mm}$f_{0}$}}
\psfrag{S44}[l][l][.7]{\raisebox{-1mm}{\hspace{-0.2mm}$f_{0}$}}
\psfrag{S40}[l][l][.7]{\raisebox{0mm}{\hspace{-1.5mm}$f_{2,0}$}}
\psfrag{S41}[l][l][.7]{\raisebox{0mm}{\hspace{-1.3mm}$f_{1,0}$}}
\psfrag{S42}[l][l][.7]{\raisebox{-0.3mm}{\hspace{-2.2mm}$f_{K,0}$}}
\psfrag{S49}[l][l][.7]{\raisebox{1mm}{\hspace{-2.5mm}$f_{K,0}$}}
\psfrag{S48}[l][l][.69]{\raisebox{0mm}{\hspace{-3mm}\textcolor{hl}{$m_{K \rightarrow 0}^{(p)}$}}}
\psfrag{S47}[l][l][.69]{\raisebox{1mm}{\hspace{-0.3mm}\textcolor{hl}{$m_{1 \rightarrow 0}^{(p)}$}}}
\psfrag{S50}[l][l][.69]{\raisebox{0mm}{\hspace{-2.3mm}\textcolor{hl}{$b_K^{(p-1)}$}}}
\psfrag{S51}[l][l][.69]{\raisebox{4.5mm}{\hspace{-2mm}\textcolor{hl}{$b_1^{(p-1)}$}}}
\psfrag{S52}[l][l][.69]{\raisebox{3mm}{\hspace{-4mm}\textcolor{hl}{$m_{k \rightarrow 0}^{(p)}$}}}
\psfrag{S53}[l][l][.69]{\raisebox{3mm}{\hspace{-1mm}\textcolor{hl}{$b_{0}^{(p-1)}$}}}
\psfrag{S54}[l][l][.69]{\raisebox{2mm}{\hspace{-2.5mm}\textcolor{hl}{$m_{0 \rightarrow 1}^{(p)}$}}}
\psfrag{S55}[l][l][.69]{\raisebox{2mm}{\hspace{-2.5mm}\textcolor{hl}{$m_{\rightarrow n}$}}}
\psfrag{S56}[l][l][.69]{\raisebox{-5mm}{\hspace{-2mm}\textcolor{hl}{$b_0^{(P)}$}}}
\includegraphics[scale=0.33]{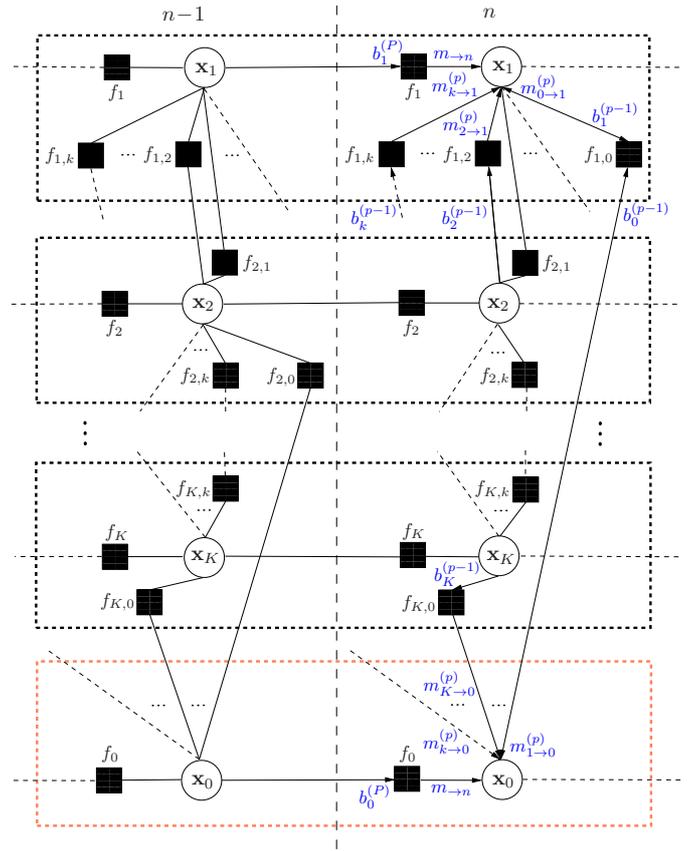}
\vspace*{-5.7mm}
\renewcommand{\baselinestretch}{1}\small\normalsize
\caption{Factor graph for CoSLAT, with sensors $k \!\in\! \{1,\dots,K\}$ and a target
($k \rmv\in\rmv 0$). 
We use the short notation $f_{k} \!\triangleq\! f(\mathbf{x}_{k,n'}|\mathbf{x}_{k,n'-1})$ and $f_{k,l} \!\triangleq\! f(y_{k,l;n'} | \mathbf{x}_{k,n'},\mathbf{x}_{l,n'})$, 
for $n' \!\in\! \{1, \dots, n\}$.
%% The part of the factor graph that is shown corresponds to one time step $n$. 
The upper 
%% three 
(black) dotted boxes correspond to the CSL part; the bottom (red) dotted box corresponds to the DTT part. All time indices are omitted for simplicity.  Only the messages and approximate marginal posteriors involved in calculating $b_{1,n}(\mathbf{x}_{1,n})$ and $b_{0,n}(\mathbf{x}_{0,n})$ are shown. Edges between black dotted boxes imply communication.}
\label{fig:coslat_fg}
\vspace*{0mm}
\end{figure}

The marginal posterior pdf $f(\mathbf{x}_{k,n}|\mathcal{Y}_{1:n})$ involved in \eqref{eq:mmse_coslat} can be calculated by
marginalization of the joint posterior pdf $f(\mathcal{X}_{0:n}|\mathcal{Y}_{1:n})$ 
of the past and present states of all sensors and the target, $\mathcal{X}_{0:n} \triangleq {\{ \mathbf{x}_{k,n'} \}}_{k \in \mathcal{A}, \, n'\in \{0,\ldots,n\}}$.
By using Bayes' rule and common assumptions \cite{wymeersch}, one can 
%% readily 
show that this joint posterior pdf factorizes as follows:
\begin{align}
f(\mathcal{X}_{0:n}|\mathcal{Y}_{1:n}) &\,\propto\, \bigg[ \prod_{k \in \mathcal{A}} \rmv f(\mathbf{x}_{k,0}) \bigg] 
  \prod_{n' \rmv= 1}^n \rmv\bigg[ \prod_{k' \rmv\in \mathcal{A}} \rmv\rmv  f(\mathbf{x}_{k'\rmv\rmv,n'}|\mathbf{x}_{k'\rmv\rmv,n'-1})  \nonumber\\
&\qquad\;\times\!\rmv \prod_{l\in \mathcal{M}_{k'\!,n'}} \!\! f(y_{k'\rmv\rmv,l;n'}|\mathbf{x}_{k'\rmv\rmv,n'} ,\mathbf{x}_{l,n'}) \bigg] \ist.
\label{eq:factoriz} \\[-6mm]
&\nonumber
\end{align}
%% The factor graph \cite{loeliger} corresponding to this factorization of $f(\mathcal{X}_{1:n}|\mathcal{Y}_{1:n})$ 
%% %% this joint posterior pdf 
%% is depicted in Fig. \ref{fig:coslat_fg}. 
%% While 
%% direct 
Calculating 
%% the marginal posterior pdf 
$f(\mathbf{x}_{k,n}|\mathcal{Y}_{1:n})$
%% , $k \in \mathcal{S}$ 
by straightforward marginaliza\-tion 
%% of $f(\mathcal{X}_{0:n}|\mathcal{Y}_{1:n})$ 
is\linebreak %%%%%%%% 
infeasible. However,
%%  in practice. However, 
an approximation of the marginal posterior, $b_{k,n}(\mathbf{x}_{k,n}) \approx f(\mathbf{x}_{k,n}|\mathcal{Y}_{1:n})$, 
%% $k \!\in\! \mathcal{A}$, 
can be obtained by executing iterative belief propagation message passing \cite{kschischang} on the factor graph corresponding to the factorization 
%% of $f(\mathcal{X}_{1:n}|\mathcal{Y}_{1:n})$ in 
\eqref{eq:factoriz}, which is shown in Fig.\ \ref{fig:coslat_fg}.
%% Because the factor graph 
%% of the CoSLAT problem in Fig. \ref{fig:coslat_fg} 
%% has loops, the BP algorithm provides only an approximate marginalization \cite{kschischang}. 
At each time $n$, $P$ message passing iterations are performed. 
%%Using the belief propagation message passing rules \cite{kschischang} with the restriction that messages are sent only forward in time
Extending the belief propagation message passing scheme for distributed CSL proposed in \cite{wymeersch} to include a noncooperative target, 
the iterated 
%% version of the 
approximate marginal posterior (AMP) of sensor or target node 
$k \!\in\! \mathcal{A}$ at 
%% time $n$ and 
message passing iter\-ation $p$, $b^{(p)}_{k,n}(\mathbf{x}_{k,n})$, is 
%% \pagebreak %%%%%%%
obtained as 
\begin{align}
&   \rmv\rmv\rmv b^{(p)}_{k,n}(\mathbf{x}_{k,n}) \ist\propto \begin{cases} 
     m_{\rightarrow n}(\mathbf{x}_{k,n}) \!\prod\limits_{\,l \in \mathcal{M}_{k,n}} \!\! m^{(p)}_{l \rightarrow k}(\mathbf{x}_{k,n}) \ist\ist, & k \!\in\! \mathcal{A}_{\sim 0}\\[4mm]
     m_{\rightarrow n}(\mathbf{x}_{0,n}) \!\prod\limits_{\,l \in \mathcal{T}_{n}} \! m^{(p)}_{l \rightarrow 0}(\mathbf{x}_{0,n}) \ist\ist, & k \!=\! 0 \,,
   \end{cases} \nonumber\\[-3mm]
\label{eq:marg}\\[-8mm]
\nonumber
\end{align}
with the ``prediction message''
\begin{equation}
\label{eq:pred_mess}
m_{\rightarrow n}(\mathbf{x}_{k,n}) \,\triangleq \int \rmv f(\mathbf{x}_{k,n}|\mathbf{x}_{k,n-1}) \, b^{(P)}_{k,n-1}(\mathbf{x}_{k,n-1}) \, d\mathbf{x}_{k,n-1} 
\vspace*{-.5mm}
\end{equation}
%% \pagebreak %%%%%%%
and the ``measurement 
%% \vspace*{1mm}
messages''      
\begin{equation}
\label{eq:ia_messcons}
m_{l \rightarrow k}^{(p)}(\mathbf{x}_{k,n}) \,\triangleq\ist \begin{cases} 
\begin{array}[t]{lr} \hspace{-2mm}\int \rmv f(y_{k,l;n} | \mathbf{x}_{k,n}\ist,\mathbf{x}_{l,n}) \, b^{(p-1)}_{l,n}(\mathbf{x}_{l,n}) \, d\mathbf{x}_{l,n} \,, \\
  \hspace{34mm} k \!\in\! \mathcal{A}_{\sim 0},\hspace{2mm}l \!\in\! \mathcal{A}_{\sim 0} \vspace{1mm}\end{array}\\[4mm]
  \begin{array}[t]{lr} \hspace{-2mm}\int \rmv f(y_{k,0;n} | \mathbf{x}_{k,n}\ist,\mathbf{x}_{0,n}) \, n^{(p-1)}_{0 \rightarrow k}(\mathbf{x}_{0,n}) \, d\mathbf{x}_{0,n} \,, \\
   \hspace{38mm} k \!\in\! \mathcal{A}_{\sim 0},\hspace{2mm}l \!=\! 0 \vspace{1mm}\end{array}\\[4mm]
\begin{array}[t]{lr}\hspace{-2mm}\int \rmv f(y_{l,0;n} | \mathbf{x}_{0,n}\ist,\mathbf{x}_{l,n}) \, n^{(p-1)}_{l \rightarrow 0}(\mathbf{x}_{l,n}) \, d\mathbf{x}_{l,n}\,,\\  
  \hspace{37mm} k \!=\! 0,\hspace{2mm}l \!\in\! \mathcal{A}_{\sim 0} \,,\vspace{1mm}\end{array}
\end{cases}
%% \vspace*{.5mm}
\end{equation}
where $n^{(p-1)}_{l \rightarrow 0}(\mathbf{x}_{l,n})$ and $n^{(p-1)}_{0 \rightarrow k}(\mathbf{x}_{0,n})$ (constituting the ``extrinsic information'') are given 
\vspace{.5mm}
by 
\begin{align}
n^{(p-1)}_{l \rightarrow 0}(\mathbf{x}_{l,n}) &\eq
     m_{\rightarrow n}(\mathbf{x}_{l,n}) \!\!\prod\limits_{k' \in \mathcal{M}_{l,n} \rmv\backslash \{0\}} \!\! m^{(p-1)}_{k' \rightarrow l}(\mathbf{x}_{l,n}) \nonumber \\[-1.5mm]
&\label{eq:ext}\\[-1.5mm]
n^{(p-1)}_{0 \rightarrow k}(\mathbf{x}_{0,n}) &\eq
     m_{\rightarrow n}(\mathbf{x}_{0,n}) \!\!\prod\limits_{k' \in \mathcal{T}_{n} \rmv\backslash \{k\}} \!\! m^{(p-1)}_{k' \rightarrow 0}(\mathbf{x}_{0,n}) \ist\ist.
       \nonumber\\[-5mm]
&\nonumber
\end{align}
However, in the proposed CoSLAT algorithm, we modify \eqref{eq:ia_messcons} in that we approximate the extrinsic information by the corresponding AMP. 
This leads to the following approximation for the measurement messages:
\begin{equation}
\label{eq:ia_mess}
m_{l \rightarrow k}^{(p)}(\mathbf{x}_{k,n}) \,\approx \begin{cases} \begin{array}[t]{lr} \hspace{-2mm}\int \rmv f(y_{k,l;n} | \mathbf{x}_{k,n}\ist,\mathbf{x}_{l,n}) \, b^{(p-1)}_{l,n}(\mathbf{x}_{l,n}) \, d\mathbf{x}_{l,n} \,, \\ \hspace{49mm} k \!\in\! \mathcal{A}_{\sim 0} \vspace{1mm}\end{array}\\[3mm]
\begin{array}[t]{lr}\hspace{-2mm}\int \rmv f(y_{l,0;n} | \mathbf{x}_{0,n}\ist,\mathbf{x}_{l,n}) \, b^{(p-1)}_{l,n}(\mathbf{x}_{l,n}) \, d\mathbf{x}_{l,n} \,,\\  
  \hspace{49mm} k \!=\! 0 \,,\end{array}
\end{cases}
\vspace*{.5mm}
\end{equation}
for all $l \!\in\! \mathcal{A}$. In this way, the costly 
%% additional 
calculation of the extrinsic information \eqref{eq:ext} is avoided. Numerical analysis showed that 
although this approximation leads to slightly overconfident AMPs, 
the estimation performance is not affected.

Because according to \eqref{eq:mess},
%% in our measurement model 
$y_{k,l;n}$ depends only on the locations of (sensor or target) nodes $k$ and $l$, 
%% the measurement message 
$m_{l \rightarrow k}^{(p)}(\mathbf{x}_{k,n})$ is 
%% always 
2D regardless of the dimension of 
%% the state 
$\mathbf{x}_{k,n}$. 
The messages and AMPs 
%% necessary for the calculation of 
needed for calculating $b^{(p)}_{1,n}(\mathbf{x}_{1,n})$ and $b^{(p)}_{0,n}(\mathbf{x}_{0,n})$ according to \eqref{eq:marg}, \eqref{eq:pred_mess}, 
and \eqref{eq:ia_mess} are depicted in Fig.\ \ref{fig:coslat_fg}. 
%% In this scheme, 
%% for algorithmic simplicity and to keep the latency low, 
Messages are sent only forward in time, and iterative message passing is performed at each time step individually \cite{wymeersch}. 
We do not send messages backward in time because this
%% is infeasible because 
would cause the computation, communication, and memory requirements as well as the latency to grow linearly with time.  
As a consequence,
%%  of this reduced message passing scheme is that 
%% at time $n$, 
%% the prediction message 
$m_{\rightarrow n}(\mathbf{x}_{k,n})$ in \eqref{eq:pred_mess} remains unchanged during the message passing iterations.
%%  at time $n$. 

%% Furthermore, we note 
The computation of the AMPs $b^{(p)}_{k,n}(\mathbf{x}_{k,n})$
%% approximate posteriors of 
%% for a sensor or target state 
according to \eqref{eq:marg} differs from pure CSL and pure DTT. 
For $k \rmv=\rmv 0$ (target), the local likelihood functions used in DTT \cite{hlinka} are replaced by the measurement messages \eqref{eq:ia_mess}. 
In this way, 
%% also 
the uncertainties about the 
%% position information 
locations of all sensors involved in DTT, $k' \!\in\rmv \mathcal{T}_n$, are taken into account. 
%% calculation of the approximate marginal posterior of the state of sensors involved in DTT 
For $k \!\in\! \mathcal{T}_n$ (a sensor involved in DTT),
also messages from the target node are considered, i.e., probabilistic information about the target location is used
by the sensors for improved self-localization.
%% These differences explain 
This probabilistic information transfer between the CSL and DTT parts is key to the superior performance of CoSLAT.
%%  compared to separate CSL and DTT (see Section \ref{sec:simres}).

%% \vspace*{-1mm}

%%%%%%%%%%%%%%%%%%%%%%%%%%%%%%%%%%%
\section{A Distributed CoSLAT Algorithm}\label{sec:slat_practical}  %% CoSLAT, Part II:\, 
%%%%%%%%%%%%%%%%%%%%%%%%%%%%%%%%%%%

\vspace*{.8mm}

Next, we develop the message passing scheme \eqref{eq:marg}, \eqref{eq:pred_mess}, 
and \eqref{eq:ia_mess} into a 
%% practical 
distributed CoSLAT algorithm.

%% describe two modifications by which the message passing scheme \eqref{eq:marg}--\eqref{eq:ia_mess}
%% can be converted into a practical distributed CoSLAT algorithm.
%% The message passing scheme \eqref{eq:marg}--\eqref{eq:ia_mess} cannot be directly implemented in a distributed manner.
%% In this section, we explain the reasons and discuss two modifications that result in a practical distributed CoSLAT algorithm.

\vspace*{-1mm}

\subsection{Nonparametric Belief Propagation}\label{sec:nbp}
%%%%%%%%%%%%%%%%%%%%%%%%%%%%%%%%%%%

\vspace*{.5mm}

Because direct calculation of 
%% the integrals in 
\eqref{eq:marg}, \eqref{eq:pred_mess}, and \eqref{eq:ia_mess} is still 
%% computationally 
infeasible, we use an
%% feasible 
%% approximation provided by 
approximate implementation 
%% of \eqref{eq:marg}--\eqref{eq:ia_mess} 
via NBP \cite{ihler, lien}. 
%% which will also be employed in our CoSLAT algorithm. 
In NBP, all AMPs and messages are represented by particles $\mathbf{x}^{(j)}$ and weights $w^{(j)}\rmv$, for $j \!\in\! \{1,\ldots,J\}$.
This particle representation is also suited to
%% capable of describing 
\pagebreak %%%%%%%%
multimodal AMPs and messages. 
NBP can be viewed 
as an extension of particle filtering to factor graphs with loops. 
%% It is able to describe also complex (multimodal) marginal posteriors and messages. 
In a CSL scenario, it
%% nonparametric BP 
%% is known to 
exhibits fast convergence and high accuracy \cite{wymeersch}.
An algorithmic description of NBP for
%% in the context of 
CSL can be found in \cite{ihler, lien}; the extension to our CoSLAT setting is straightforward.
%%  and will not be discussed here. 
In the CoSLAT message passing scheme,
%%  discussed in this paper 
all particles representing a message have equal weights, i.e., $w^{(j)} \!\equiv\! 1/J$.
%%  for all $j \in \{1,\dots,J\}$. 

In addition to the particle representation of messages,
%%  mentioned above, 
NBP uses 
%% also 
an approximate kernel representation that can be easily derived from the particle representation.
This kernel representation provides a closed-form expression 
%% of the messages, which 
that can be evaluated at any given point. 
This is necessary for performing
%% the nonparametric BP implementation of 
the message multiplication in \eqref{eq:marg} and 
%% will also be necessary
for using the LC (see Section \ref{sec:lc}). Given a set of particles and weights $\big\{ (\mathbf{x}^{(j)},w^{(j)}) \big\}_{j=1}^J$ 
representing a measurement message $m(\mathbf{x})$, the kernel representation
%% density estimate 
of $m(\mathbf{x})$ is obtained as
\begin{equation}
\label{eq:kermess}
\hat{m}(\mathbf{x}) \,=\ist \sum_{j = 1}^J w^{(j)} K(\tilde{\mathbf{x}} \!-\! \tilde{\mathbf{x}}^{(j)}) \,,
\end{equation}
%% Here, the \emph{kernel} $K(\mathbf{x})$ is a symmetric, unimodal function whose width can be tuned.
where, as before, the 2D vector $\tilde{\mathbf{x}}$ denotes the location part of the state $\mathbf{x}$. 
A standard choice for the kernel $K(\tilde{\mathbf{x}})$ in the 2D localization scenario is the 2D 
%% two-dimensional circularly symmetric 
Gaussian function
%%  given by 
$K(\tilde{\mathbf{x}}) = (2 \pi \sigma_{\!K}^2)^{-1} \exp\rmv \big(\!\rmv -\rmv\rmv \| \tilde{\mathbf{x}} \|^2/(2 \sigma_{\!K}^2) \big)$.
%% \begin{equation}
%% K_{\sigma_k}(\mathbf{x}) = \frac{1}{2 \pi \sigma_k^2} \exp \bigg(\hspace{-1mm}-\rmv\frac{\| \mathbf{x} \|^2}{2 \sigma_k^2} \bigg).
%% \end{equation}
The variance
%% width parameter 
%% bandwidth 
$\sigma_{\!K}^2$ is usually estimated from the particles and weights.
When $\sigma_{\!K}^2$ is large, $\hat{m}(\mathbf{x})$ is smooth but some of the finer details of $m(\mathbf{x})$ may be smoothed out;
when $\sigma_{\!K}^2$ is small, $\hat{m}(\mathbf{x})$ preserves more of these fine details but may exhibit some artificial structure not present in 
$m(\mathbf{x})$ \cite{lien, botev07}. 

\vspace*{-1.5mm}

%% \newpage %%%%%%%

\subsection{Likelihood Consensus Based Computation of $b^{(p)}_{0,n}(\mathbf{x}_{0,n})$}\label{sec:lc}
%%%%%%%%%%%%%%%%%%%%%%%%%%%%%%%%%%%

\vspace*{1mm}

In CSL, the NBP message passing scheme 
%% described above 
can be performed in a 
%% fully 
distributed manner 
using only local intersensor communications. 
With CoSLAT, 
%% however, 
a distributed implementation 
%% with only local communications 
is complicated by the 
%% following problem.
fact that the target node is noncooperative and therefore some vital information is not communicated to the sensors.
%% scheme for both the calculation of the approximate marginal posterior of the target and the sensor state according to 
%% \eqref{eq:marg}--\eqref{eq:ia_mess}, necessary messages are no available.
More specifically, calculating the AMP of the target state, $b^{(p)}_{0,n}(\mathbf{x}_{0,n})$, according to \eqref{eq:marg} requires
%% --\eqref{eq:ia_mess}
%% (or using the nonparametric BP implementation of \eqref{eq:marg}), 
the product of measurement messages $\prod_{l \ist\in\ist \mathcal{T}_{n}} m^{(p)}_{l \rightarrow 0}(\mathbf{x}_{0,n})$. 
%% is required.
%%  at each sensor. 
Unfortunately, this message product is not available at the sensors.

We solve this problem by using the LC scheme, which was 
%% originally 
proposed in a different context in \cite{hlinka}.
Consider a sensor $l \!\in\! \mathcal{T}_{n}$ and the 
%% ???locally available 
kernel approximation $\hat{m}^{(p)}_{l \rightarrow 0}(\mathbf{x}_{0,n})$ (see \eqref{eq:kermess}) 
of the measurement message $m^{(p)}_{l \rightarrow 0}(\mathbf{x}_{0,n})$, which was calculated at sensor $l$.
%% and is thus locally available.
Following the LC principle, the logarithm of $\hat{m}^{(p)}_{l \rightarrow 0}(\mathbf{x}_{0,n})$ is approximated by a finite-order basis expansion:
\begin{equation}
\label{eq:basis_exp}
\log \, \hat{m}^{(p)}_{l \rightarrow 0}(\mathbf{x}_{0,n}) \,\approx\, \sum_{r=1}^{R} \beta^{(p)}_{l;n,r}(y_{l,0;n}) \,\varphi_r(\mathbf{x}_{0,n}) \,.
\end{equation}
Here, the basis functions $\varphi_r(\mathbf{x}_{0,n})$ do not depend on $l$, i.e., the same set of basis functions is used by all sensors.
The expansion coefficients $\beta^{(p)}_{l;n,r}(y_{l,0;n})$, $r \in \{1,\ldots,R\}$ can be calculated locally at sensor $l$ by least squares fitting 
%% \cite{bjorck96} 
using the particles of the prediction message $m_{\rightarrow n}(\mathbf{x}_{l,n})$ as reference points (cf.\ \cite{hlinka}).
Furthermore, we formally set $\beta^{(p)}_{l;n,r}(y_{l,0;n}) \rmv=\rmv 0$ for all $r \rmv\in\rmv \{1,\dots,R\}$ if $l \rmv\notin\rmv \mathcal{T}_n$.

The local approximations \eqref{eq:basis_exp} 
%% at the various sensors $l \!\in\! \mathcal{A}_{\sim 0}$ 
entail the following approximation of the desired message product:
\begin{align}
\prod_{l \ist\in\ist \mathcal{T}_{n}} \! \hat{m}^{(p)}_{l \rightarrow 0}(\mathbf{x}_{0,n}) 
  &\,\approx \prod_{l \ist\in\ist \mathcal{T}_{n}} \! \exp\! \Bigg( \sum_{r=1}^{R} \beta^{(p)}_{l;n,r}(y_{l,0;n}) \,\varphi_r(\mathbf{x}_{0,n}) \rmv \Bigg) \nonumber\\[.5mm]
  &\,=\, \exp\! \Bigg( \sum_{r=1}^{R} B^{(p)}_{n,r} \,\varphi_r(\mathbf{x}_{0,n}) \rmv \Bigg) \,, \label{eq:M_prod}
\\[-7.5mm]
\nonumber
\end{align}
with
\vspace{1.5mm}
\begin{equation}
B^{(p)}_{n,r} \,\triangleq \sum_{l \in \mathcal{T}_n} \!\beta^{(p)}_{l;n,r}(y_{l,0;n}) = \sum_{l \in \mathcal{A}_{\sim 0}} \!\beta^{(p)}_{l;n,r}(y_{l,0;n}) \,,
\label{eq:b_coslat}
\vspace{-.8mm}
\end{equation}
where the last equation follows because $\beta^{(p)}_{l;n,r}(y_{l,0;n}) \!=\! 0$ for all $l \!\notin\! \mathcal{T}_n$.
%As in LC, all $B_r$ are calculated with local communication only using average consensus \cite{olfatisaber} or gossip algorithms \cite{dimakis10}. 
%% Because of the sum expression \eqref{eq:b_coslat}, 
The coefficients $B_{n,r}^{(p)}$ in \eqref{eq:b_coslat} can be computed at\linebreak %%%%%%% 
each sensor by running 
$R$ parallel instances of an average consensus algorithm or a gossip algorithm \cite{olfatisaber,dimakis10}.
%%  in the sensor network.
%%  as described in \cite{hlinka}. 
This requires only local communications between neighboring sensors.
After convergence of the consensus or gossip algorithms, an approximation of 
the functional form of 
$\prod_{l \ist\in\ist \mathcal{T}_{n}} \! \hat{m}^{(p)}_{l \rightarrow 0}(\mathbf{x}_{0,n})$
is available at each sensor.
%%  $k \!\in\! \mathcal{A}_{\sim 0}$;
%% this approximation can be evaluated at any $\mathbf{x}_{0,n}$.
%%  with only local communications.
Each sensor 
%% $k \!\in\! \mathcal{A}_{\sim 0}$ 
is then able to calculate a particle representation of 
%% the posterior pdf of the target state at time $n$ and message passing iteration $p$, 
$m_{\rightarrow n}(\mathbf{x}_{0,n}) \prod_{l \ist\in\ist \mathcal{T}_{n}} \! \hat{m}^{(p)}_{l \rightarrow 0}(\mathbf{x}_{0,n}) \approx b^{(p)}_{0,n}(\mathbf{x}_{0,n})$ 
(see \eqref{eq:marg}) based on the importance sampling principle \cite{doucet}.
More specifically, weights $\big\{ w^{(j)}_{0,n} \big\}_{j=1}^J$ associated with the 
%% equally weighted 
particles $\big\{ \mathbf{x}^{(j)}_{0,n} \big\}_{j=1}^J$ representing 
%% the prediction message 
$m_{\rightarrow n}(\mathbf{x}_{0,n})$ are obtained by evaluating the approximation \eqref{eq:M_prod} of 
%% the message product 
$\prod_{l \ist\in\ist \mathcal{T}_{n}} \! \hat{m}^{(p)}_{l \rightarrow 0}(\mathbf{x}_{0,n})$ at the 
%% particles 
$\mathbf{x}^{(j)}_{0,n}$, i.e., by calculating $w^{(j)}_{0,n} = \exp\! \Big( \!\sum_{r=1}^{R} \rmv B^{(p)}_{n,r} \,\varphi_r(\mathbf{x}^{(j)}_{0,n}) \Big)$ 
%% (see \eqref{eq:M_prod}) 
for all 
%% $\mathbf{x}^{(j)}_{0,n}$, 
$j \!\in\! \{1,\dots,J\}$. Then, 
%% procedure 
a resampling step \cite{doucet} is performed to obtain 
%% a set of 
equally weighted particles representing 
%% the marginal posterior 
$b^{(p)}_{0,n}(\mathbf{x}_{0,n})$. 

Once a particle approximation of $b^{(p)}_{0,n}(\mathbf{x}_{0,n})$ is available at each sensor, computations in the CSL part of the factor graph 
(cf.\ the upper 
%% three 
dotted boxes in Fig.\ \ref{fig:coslat_fg}) at message passing iteration $p \ist+ 1$ can be performed in a distributed way 
using NBP %% \linebreak %%%%%%%%%
 as described in \cite{ihler,lien}. Thus, each sensor $k \!\in\! \mathcal{A}_{\sim 0}$ is able to calculate 
%% the 
approximate marginals 
of its own state $\mathbf{x}_{k,n}$ and of the target state $\mathbf{x}_{0,n}$ by means of the NBP implementation of 
\eqref{eq:marg}, \eqref{eq:pred_mess}, and \eqref{eq:ia_mess}, using information that is either locally available or obtained through local communication.

\vspace*{-1mm}

%%%%%%%%%%%%%%%%%%%%%%%%%%%%%%%%%%%
\section{Simulation Results}\label{sec:simres}
%%%%%%%%%%%%%%%%%%%%%%%%%%%%%%%%%%%

\vspace*{.5mm}

\begin{figure}
\vspace{1mm}
\centering
\psfrag{s01}[t][t][0.9]{\color[rgb]{0,0,0}\setlength{\tabcolsep}{0pt}\begin{tabular}{c}\raisebox{-1mm}{$x_1$ coordinate}\end{tabular}}
\psfrag{s02}[b][b][0.9]{\color[rgb]{0,0,0}\setlength{\tabcolsep}{0pt}\begin{tabular}{c}$x_2$ coordinate\end{tabular}}
\psfrag{x01}[t][t][0.7]{$-30$}
\psfrag{x02}[t][t][0.7]{$-20$}
\psfrag{x03}[t][t][0.7]{$-10$}
\psfrag{x04}[t][t][0.7]{$0$}
\psfrag{x05}[t][t][0.7]{$10$}
\psfrag{x06}[t][t][0.7]{$20$}
\psfrag{x07}[t][t][0.7]{$30$}
\psfrag{x08}[t][t][0.7]{$40$}
\psfrag{x09}[t][t][0.7]{$50$}
\psfrag{x10}[t][t][0.7]{$60$}
\psfrag{v01}[r][r][0.7]{$-30$}
\psfrag{v02}[r][r][0.7]{$-20$}
\psfrag{v03}[r][r][0.7]{$-10$}
\psfrag{v04}[r][r][0.7]{$0$}
\psfrag{v05}[r][r][0.7]{$10$}
\psfrag{v06}[r][r][0.7]{$20$}
\psfrag{v07}[r][r][0.7]{$30$}
\psfrag{v08}[r][r][0.7]{$40$}
\psfrag{v09}[r][r][0.7]{$50$}
\psfrag{v10}[r][r][0.7]{$60$}
\includegraphics[scale=0.43]{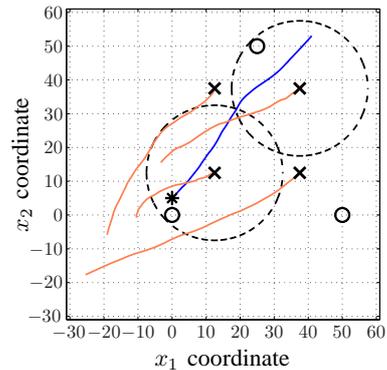}
\vspace{1.6mm}
\renewcommand{\baselinestretch}{1}\small\normalsize
\caption{Network topology used for the simulations, along with a realization of the target and sensor trajectories. Initial mobile sensor locations 
are indicated by crosses, anchor locations by circles, and the initial target location by a star. The big dashed 
%% black 
circles indicate the measurement regions of the upper-right sensor (in both scenarios) and of the lower-left sensor (in scenario 2).}
\label{fig:top}
\end{figure}

We consider a 
%% sensor 
network 
%% consisting 
of $K\!=\!7$ sensors, of which four are mobile sensors and three are anchors (i.e., 
static sensors with perfect location information modeled via Dirac-shaped priors).
%%  at their true location and velocity zero). 
The sensors are placed within a field 
of size 50$\ist\times\ist$50. Each sensor has a communication range of 56 and localizes itself and the target.
We consider two 
%% different 
scenarios.
In scenario 2, which is shown in Fig. \ref{fig:top}, the upper-right and lower-left sensors have a measurement radius of 
%% $D_k \!=\! 20\ist$m 
20, and therefore, initially (at time $n\!=\!0$), they do not have enough partners for self-localization. With conventional CSL, at $n\!=\!0$, these  
sensors have a multimodal marginal posterior and are thus unable to localize themselves.
%% determine their locations. 
%% However, they are able to communicate with ???the other sensors in the network. 
The measurement regions of the other five sensors cover the entire field.  
Scenario 1 
%% (not shown) 
differs from scenario 2 in that also the lower-left sensor covers the entire field. 
%% In both scenarios, each sensor communication range of ???m.

\begin{figure*}[t!]
\centering\hspace*{7mm}
\begin{minipage}[H!]{0.40\textwidth}
\psfrag{s01}[t][t][0.9]{\color[rgb]{0,0,0}\setlength{\tabcolsep}{0pt}\begin{tabular}{c}\raisebox{-.7mm}{$n$}\end{tabular}}
\psfrag{s02}[b][b][0.7]{\color[rgb]{0,0,0}\setlength{\tabcolsep}{0pt}\begin{tabular}{c}\vspace{0.0cm}{\large RMSE [m]}\end{tabular}}
\psfrag{s05}[l][l]{\color[rgb]{0,0,0}Tracking RMSE of CoSLAT}
\psfrag{s06}[l][l][0.67]{\color[rgb]{0,0,0}Self-localization RMSE of CSL \cite{lien} + DTT \cite{hlinka}}
\psfrag{s07}[l][l][0.67]{\color[rgb]{0,0,0}\raisebox{0.5mm}{Self-localization RMSE of CoSLAT}}
\psfrag{s08}[l][l][0.67]{\color[rgb]{0,0,0}Tracking RMSE of CSL \cite{lien} + DTT \cite{hlinka}}
\psfrag{s09}[l][l][0.67]{\color[rgb]{0,0,0}Tracking RMSE of CoSLAT}
\psfrag{s13}[l][l][1]{\raisebox{-27mm}{\hspace{27.8mm}{\small (a)}}}
\psfrag{s11}[][]{\color[rgb]{0,0,0}\setlength{\tabcolsep}{0pt}\begin{tabular}{c} \end{tabular}}
\psfrag{s12}[][]{\color[rgb]{0,0,0}\setlength{\tabcolsep}{0pt}\begin{tabular}{c} \end{tabular}}
\psfrag{x01}[t][t][0.67]{$10$}
\psfrag{x02}[t][t][0.67]{$20$}
\psfrag{x03}[t][t][0.67]{$30$}
\psfrag{x04}[t][t][0.67]{$40$}
\psfrag{x05}[t][t][0.67]{$50$}
\psfrag{x06}[t][t][0.67]{$60$}
\psfrag{x07}[t][t][0.67]{$70$}
\psfrag{v01}[r][r][0.67]{$0\!$}
\psfrag{v02}[r][r][0.67]{$2$}
\psfrag{v03}[r][r][0.67]{$4$}
\psfrag{v04}[r][r][0.67]{$6$}
\psfrag{v05}[r][r][0.67]{$8$}
\psfrag{v06}[r][r][0.67]{$10$}
\psfrag{v07}[r][r][0.67]{$12$}
\psfrag{v08}[r][r][0.67]{$14$}
\psfrag{v09}[r][r][0.67]{$16$}
\psfrag{v10}[r][r][0.67]{$18$}
\psfrag{v11}[r][r][0.67]{$20\!$}
\includegraphics[height=46mm, width=67mm]{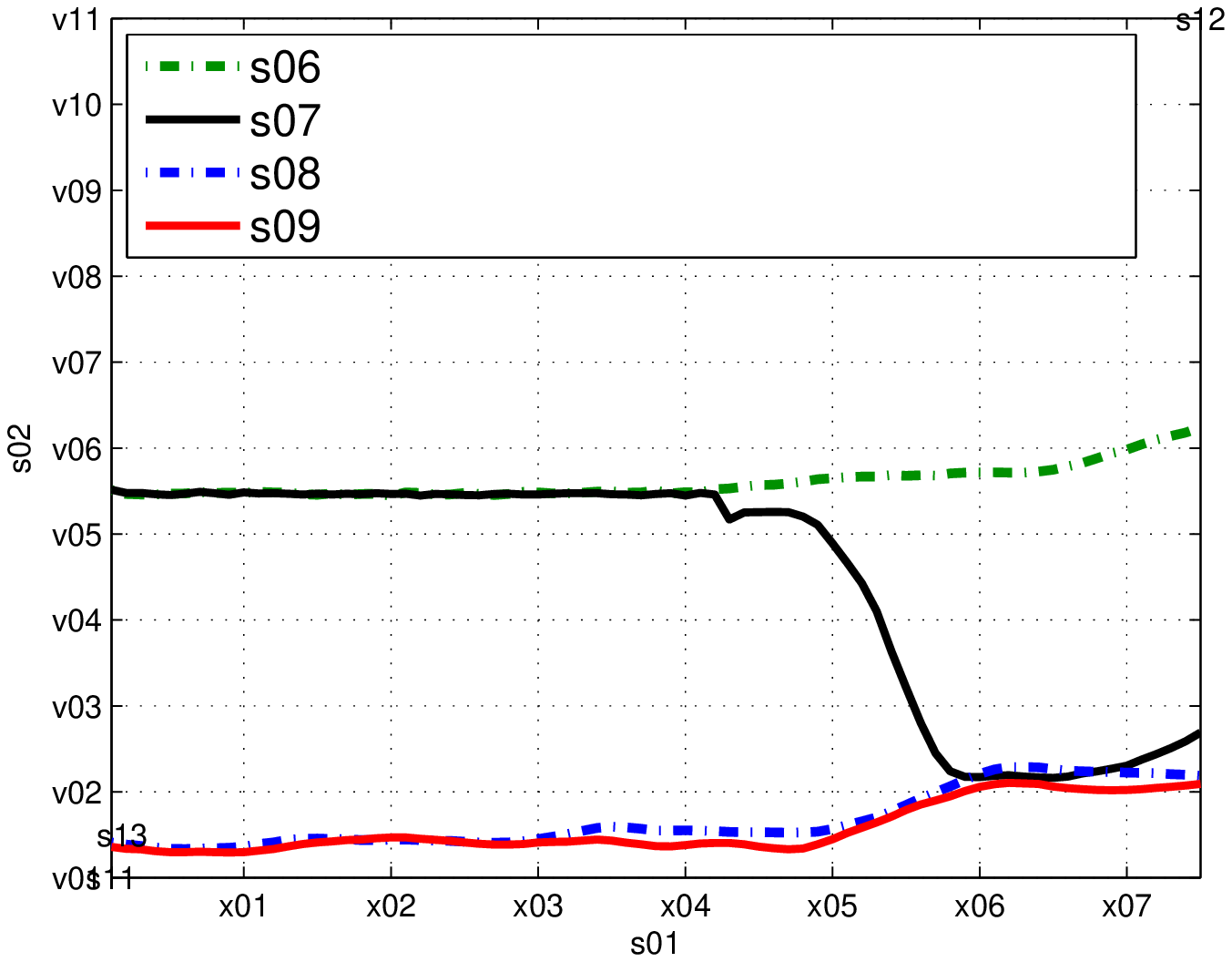}
\end{minipage}\hspace{10mm}
%%%%%%%%%%%%%%%%%%%%%%%%%%%%%%%%
\begin{minipage}[H!]{0.40\textwidth}
\psfrag{s01}[t][t][0.9]{\color[rgb]{0,0,0}\setlength{\tabcolsep}{0pt}\begin{tabular}{c}\raisebox{-.7mm}{$n$}\end{tabular}}
\psfrag{s02}[b][b][0.7]{\color[rgb]{0,0,0}\setlength{\tabcolsep}{0pt}\begin{tabular}{c}\vspace{0.0cm}{\large RMSE [m]}\end{tabular}}
\psfrag{s05}[l][l]{\color[rgb]{0,0,0}Tracking RMSE of CoSLAT}
\psfrag{s06}[l][l][0.67]{\color[rgb]{0,0,0}Self-localization RMSE of CSL \cite{lien} + DTT \cite{hlinka}}
\psfrag{s07}[l][l][0.67]{\color[rgb]{0,0,0}\raisebox{0.5mm}{Self-localization RMSE of CoSLAT}}
\psfrag{s08}[l][l][0.67]{\color[rgb]{0,0,0}Tracking RMSE of CSL \cite{lien} + DTT \cite{hlinka}}
\psfrag{s09}[l][l][0.67]{\color[rgb]{0,0,0}Tracking RMSE of CoSLAT}
\psfrag{s13}[l][l][1]{\raisebox{-27mm}{\hspace{27.8mm}{\small (b)}}}
\psfrag{s11}[][]{\color[rgb]{0,0,0}\setlength{\tabcolsep}{0pt}\begin{tabular}{c} \end{tabular}}
\psfrag{s12}[][]{\color[rgb]{0,0,0}\setlength{\tabcolsep}{0pt}\begin{tabular}{c} \end{tabular}}
\psfrag{x01}[t][t][0.67]{$10$}
\psfrag{x02}[t][t][0.67]{$20$}
\psfrag{x03}[t][t][0.67]{$30$}
\psfrag{x04}[t][t][0.67]{$40$}
\psfrag{x05}[t][t][0.67]{$50$}
\psfrag{x06}[t][t][0.67]{$60$}
\psfrag{x07}[t][t][0.67]{$70$}
\psfrag{v01}[r][r][0.67]{$0\!$}
\psfrag{v02}[r][r][0.67]{$2$}
\psfrag{v03}[r][r][0.67]{$4$}
\psfrag{v04}[r][r][0.67]{$6$}
\psfrag{v05}[r][r][0.67]{$8$}
\psfrag{v06}[r][r][0.67]{$10$}
\psfrag{v07}[r][r][0.67]{$12$}
\psfrag{v08}[r][r][0.67]{$14$}
\psfrag{v09}[r][r][0.67]{$16$}
\psfrag{v10}[r][r][0.67]{$18$}
\psfrag{v11}[r][r][0.67]{$20\!$}
\includegraphics[height=46mm, width=67mm]{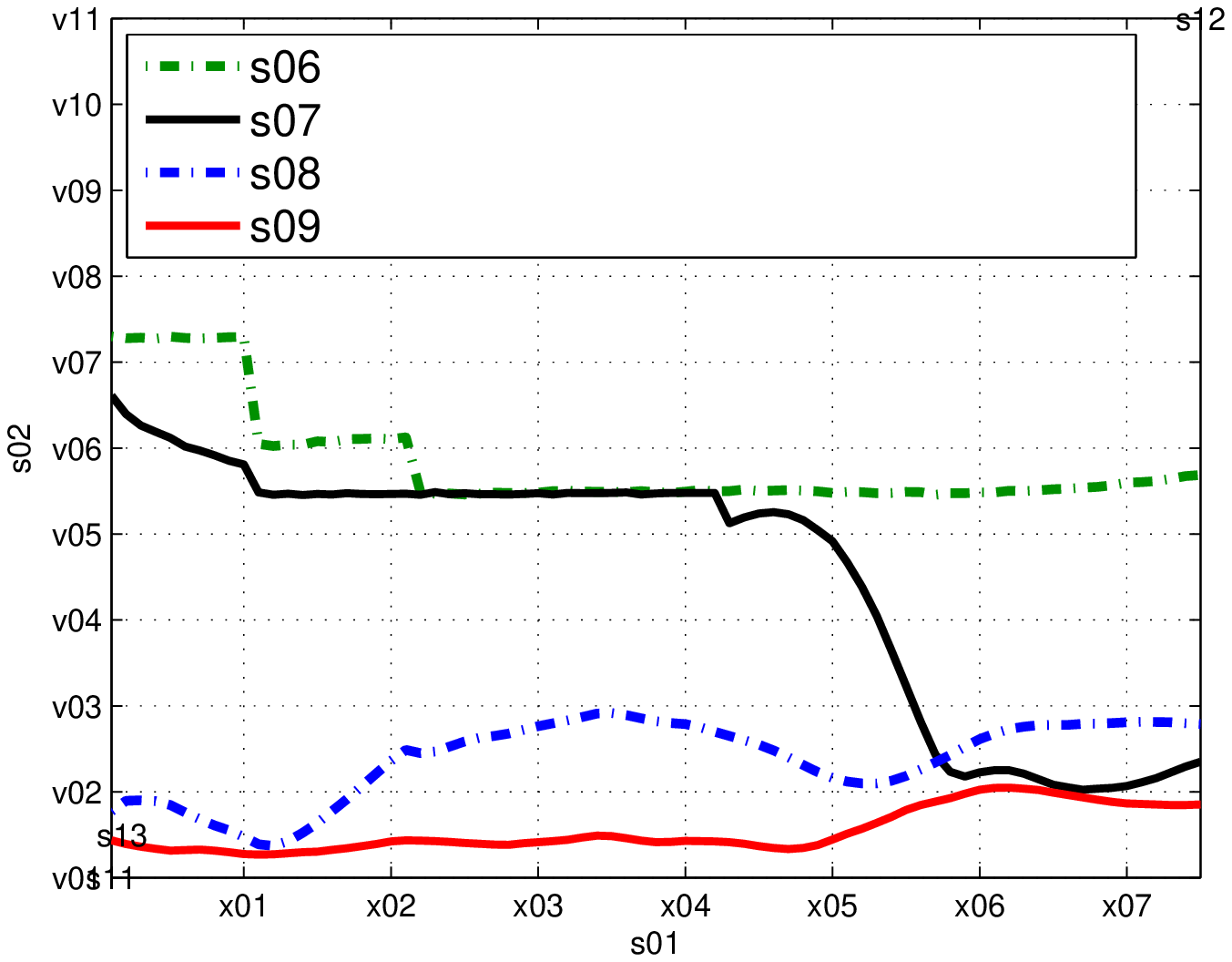}
\end{minipage}
\vspace{9mm}
\renewcommand{\baselinestretch}{1.2}\small\normalsize
\caption{Average root-mean-square errors (RMSE) of sensor self-localization and target tracking versus time $n$, for 
(a) scenario 1 and (b) scenario 2.}
\label{fig:plot}
\vspace{0mm}
\end{figure*}

The states of the mobile sensors and the target consist of location and velocity, i.e.,
$\mathbf{x}_{k,n} = [x_{1,k,n}\ist\ist\ist\ist x_{2,k,n}\ist\ist\ist\ist \dot{x}_{1,k,n}\ist\ist\ist\ist \dot{x}_{2,k,n}]^\text{T}\rmv$. 
Mobile sensor trajectories are created by using a Dirac-shaped location prior at the locations indicated in Fig.\ \ref{fig:top}. 
However, in the simulations of the algorithms, all mobile sensors have a location prior that is uniform on $[-500,500]\times[-500,500]$.
Furthermore, we used a Gaussian sensor velocity prior with mean $\bm{\mu}_{k,0} = [-0.1\hspace{1.2mm}-0.1]^\text{T}$ 
and covariance matrix $\mathbf{C}_{k,0} = \mathrm{diag}\{0.1, 0.1\}$ and a Gaussian target state prior with mean 
$\bm{\mu}_{0,0} = [0\hspace{1.6mm}5\hspace{1.6mm}0.4\hspace{1.6mm}0.4]^\text{T}$ and covariance matrix $\mathbf{C}_{0,0} = \mathrm{diag}\{1, 1, 0.001, 0.001\}$.
The mobile sensors and the target evolve independently according to 
$\mathbf{x}_{k,n} = \mathbf{G}\mathbf{x}_{k,n-1} + \mathbf{W}\mathbf{u}_{k,n}$, $n \!=\! 1,2,\dots$ \cite{rong},
where the matrices $\mathbf{G} \!\in\! \mathbb{R}^{4\times 4}$ and $\mathbf{W} \rmv\!\in\! \mathbb{R}^{4\times 2}$ are chosen as in \cite{hlinka} 
and the driving noise vectors $\mathbf{u}_{k,n} \!\in\! \mathbb{R}^2$ are Gaussian, i.e.,
$\mathbf{u}_{k,n} \!\sim\! \mathcal{N}(\mathbf{0},\sigma_u^2\mathbf{I})$, with variance $\sigma_u^2 \!=\! 0.0005$ and 
with $\mathbf{u}_{k,n}$ and $\mathbf{u}_{k'\!,n'}$  independent unless
%% \linebreak %%%%%%%% 
$(k,n) \!=\! (k'\!,n')$. 
We performed 500 simulation runs. In each
%% \linebreak %%%%%%%%
run, the sensors and the target move along the specific 
%% \nolinebreak %%%%%%%% 
trajec\-tory realizations shown in Fig.\ \ref{fig:top}.
%% whereas the observation noise and the ???initial particles are chosen randomly.
%% Mobile sensors and the target move along the red trajectories shown in Fig. \ref{fig:top}. 
The observation noise variance is $\sigma_v^2 = 2$.
%%  for all sensors. 
Each mobile sensor starts moving only when it is sufficiently localized in the sense that the sum of its estimated location
coordinate variances 
%% trace of their location covariance matrix 
is below $5\sigma_v^2$.

We compare the performance of the proposed CoSLAT algorithm with that of a state-of-the-art reference method,
%%  performing separate CSL and DTT. 
%% More specifically, the 
%% The reference method 
which separately performs CSL by means of NBP as described in
%% with a particle representation of the messages 
\cite{lien} and DTT by means of the LC-based distributed particle filter presented in \cite{hlinka}. The DTT method uses the sensor location estimates provided by 
the CSL method.
%%  instead of the true sensor locations.
In both the CoSLAT method and the reference method, the LC scheme uses an average consensus \cite{olfatisaber} with five 
%% consensus 
iterations, and the 
%% underlying
basis expansion is a third-order polynomial approximation \cite{hlinka}, resulting in an expansion order of $R \!=\! 16$.
%% ???We performed the LC scheme until full convergence (exact sum calculation). 
The NBP scheme performs $P \!=\! 3$ message passing iterations.
%%  and uses $J = 500$ particles.
The kernel variance for the measurement messages (cf.\ \eqref{eq:kermess})
%% use 2D Gaussian kernels with the variance 
is chosen as $\sigma_{\!K}^2 \!=\! \sigma_v^2$, as recommended in \cite{ihler}.
The number of particles used by both NBP and the distributed particle filter is $J \rmv= 500$.

Fig.\ \ref{fig:plot} shows the simulated 
%% average 
root-mean-square self-localiza\-tion 
%% error and mean-square 
and target localization errors 
%% versus time $n$.
for $n = 0,\ldots,75$.
%% , for 75 simulated time steps.
%%  $n$ 
These errors were determined by averaging over all sensors and all simulation runs.
%% It can be seen that 
In scenario 1, for $n \!>\! 43$, the self-localization error of CoSLAT is seen to be significantly smaller than that of the reference method.
%%  larger than 43.
%% outperforms the reference method after $n = 43$ in terms of self-localization performance. 
This 
%% can be explained as follows. With 
is because with pure CSL, the upper-right sensor has not enough partners for self-localization, 
whereas with CoSLAT, for $n \!>\! 43$, the upper-right sensor can use the measurement of its distance to the target to calculate the message from the target node, $m^{(p)}_{0 \rightarrow k}(\mathbf{x}_{k,n})$, and use this additional information to improve its self-localization performance. 
The tracking performance of CoSLAT 
in scenar-\linebreak %%%%%%%
io 1 is similar to that of the reference method.

%% \pagebreak %%%%%%%

In scenario 2, for $n \!>\! 43$,
%% where also the lower-left sensor has not enough partners for self-localization, 
the self-localization error of CoSLAT is again much smaller than that of the reference method. 
%% for the reasons explained above. 
In addition, it is also smaller for $n \!<\! 22$. This is because in scenario 2,
for $n \!<\! 22$, also the lower-left sensor has not enough partners for self-localization when pure CSL is used. 
Furthermore, the target tracking error of CoSLAT is now significantly smaller than that of the reference method for almost all times.
%% In fact, the target tracking error of the reference method is higher than in scenario 1 
This is because with separate CSL and DTT, the poor self-localization
of the lower-left sensor at $n \!<\! 22$ degrades the target tracking performance. This higher target tracking error is retained 
%% over time (
for $n \rmv\rmv\geq\! 22$ even when all sensors involved in the target tracking are well localized.
%%  by both methods.
%%For $n \!>\! 43$, the high target tracking error of the reference method is due to the fact that ???
%%By contrast, with CoSLAT, ???
%% i.e. also when all sensors involved in the target tracking are localized by both methods ($22 \leq n \leq 43$) 
%% the target localization performance of CoSLAT is increased compared to separate CSL and DTT. 

%% \newpage %%%%%%%%

\vspace{-.5mm}

%%%%%%%%%%%%%%%%%%%%%%%%%%%%%%%%%%%
\section{Conclusion}\label{sec:concl}
%%%%%%%%%%%%%%%%%%%%%%%%%%%%%%%%%%%

The novel framework of \emph{cooperative simultaneous localization and tracking} (CoSLAT) 
provides a complete and consistent combination of cooperative self-localization (CSL) and distributed target tracking (DTT).
Starting from a factor graph formulation of the CoSLAT problem, we developed a particle-based, distributed message passing algorithm for CoSLAT
that
%% combines nonparametric belief propagation with the likelihood consensus scheme.
%% It is able to outperform existing algorithms by exchanging 
performs a probabilistic information transfer between CSL and DTT.
%%  that is allowed by the CoSLAT framework. A fundamental difficulty that was surmounted in this scheme is that certain 
%% messages needed to calculate the target states are not available locally at the sensors. The proposed CoSLAT algorithm 
%% uses LC for a consensus-based distributed calculation of the lacking messages. 
Simulation results demonstrated significant improvements in both self-localization 
%% performance 
and target tracking performance 
%% achieved with the CoSLAT algorithm in comparison to 
compared to 
%% separate 
state-of-the-art 
%% CSL and DTT 
\vspace{-1mm}
algorithms.

%*******************************************************************************
\section*{Acknowledgment} 
%*******************************************************************************

%% \vspace{-.5mm}

The authors would like to thank Prof.\ Henk Wymeersch for illuminating comments.

\vspace{-.5mm}

\renewcommand{\baselinestretch}{0.89}\normalsize\footnotesize
\bibliographystyle{ieeetr_noParentheses}
\bibliography{references}

\end{document}